\documentclass[aps,pre,twocolumn,showpacs]{revtex4}
\usepackage{amssymb}
\usepackage{amsmath}
\usepackage{epsfig}
\usepackage{color}

\newcommand {\cn} {{\rm cn}}

\newcommand {\sn} {{\rm sn}}
\newcommand {\dn} {{\rm dn}}
\begin{document}

\title{Tristability in the pendula chain}
\author {Ramaz Khomeriki${}^{1,2}$,  J\'er\^ome Leon${}^3$}
\affiliation {${\ }^{(1)}$ Physics Department, Tbilisi State University, 0128
  Tbilisi (Georgia) \\ 
${\ }^{(2)}$ Max-Planck-Institut fur Physik komplexer Systeme, 01187 Dresden 
(Germany) \\
${\ }^{(3)}$ Laboratoire de Physique Th\'eorique et Astroparticules 
  CNRS-IN2P3 (UMR5207), Universit\'e Montpellier 2, 34095 Montpellier (France) }

\begin{abstract} Experiments on a chain of coupled pendula driven periodically
at one end demonstrate the existence of a novel regime which produces an output
frequency at an odd fraction of the driving frequency. The new stationary state
is then obtained on numerical simulations and modeled with an analytical
solution of the continuous sine-Gordon equation that resembles a kink-like
motion back and forth in the restricted geometry of the chain. This solution
differs from the expressions used to understand nonlinear bistability where the
\textit{synchronization constraint} was the basic assumption. As a result the
short pendula chain is shown to possess tristable stationary states and to act
as a frequency divider.
\end{abstract}

\maketitle

\section{Introduction.}

The sine-Gordon model and its discrete analogue, the Frenkel-Kontorova chain,
are among the most prominent equations of nonlinear physics, and have attracted
interest of people working in quite different fields, see e.g. books
\cite{kivsharbook,remoissenet,scott}. In particularly, topological (kinks) and
nontopological (breathers) solutions of the sine-Gordon equation describe the
dynamics of nonlinear excitations in various spatially modulated systems, e.g.
dislocations in crystals \cite{crystal}, magnetic and ferroelectric domain walls
motion \cite{wall}, vortices in arrays of Josephson junctions \cite{JJ}, etc. At
the same time the model has a simple experimental counterpart, namely the chain
of linearly coupled pendula \cite{scott} which offers an interesting opportunity
to easily visualize all the main nonlinear characteristics of the sine-Gordon
system. Then this simple laboratory tool allows to observe novel effects
\cite{pendula1,pendula2,pendula3,pendula4} which may then apply in completely
different physical situations.
 
As a matter of fact, a recent experimental discovery of supratransmission effect
in the  pendula chain \cite{supra1} has led to the study of similar phenomena in
optical Bragg gratings \cite{spire}, Josephson Junction transmission line
\cite{supra2}, waveguide arrays \cite{Leon-PRE,ramaz1}. By this approach,
many similar phenomena observed in the same systems
\cite{Winful1,Chen,Olsen,Barday,Kivshar} have been identified as effects of
\textit{nonlinear bistability}. Moreover it allowed us to predict the
existence of bistable magnetization profiles in thin magnetic films \cite{RJ1}
and to suggest ultrasensitive detectors (or digital amplifyers) in optical
waveguides \cite{RJ2} quantum Hall bilayers \cite{RJ3} and Josephson Junction
parallel arrays \cite{RJ4}. 

The bistability property can be simply formulated by saying that \cite{RJ5} a
given periodic boundary driving may produce two completely different stationary
states: one which tends to the linear evanescent profile at vanishing amplitude,
the other one which can be qualitatively understood as a portion of the
stationary breather-like solution which exists only if the system size is
comparable with the characteristic length of the fundamental (continuous)
breather solution. This is a main difference with most of the earlier studies on
sine-Gordon model where the semi-infinite chain has been examined (see e.g. Ref.
\cite{ref}). 

\begin{figure}[ht]
\epsfig{file=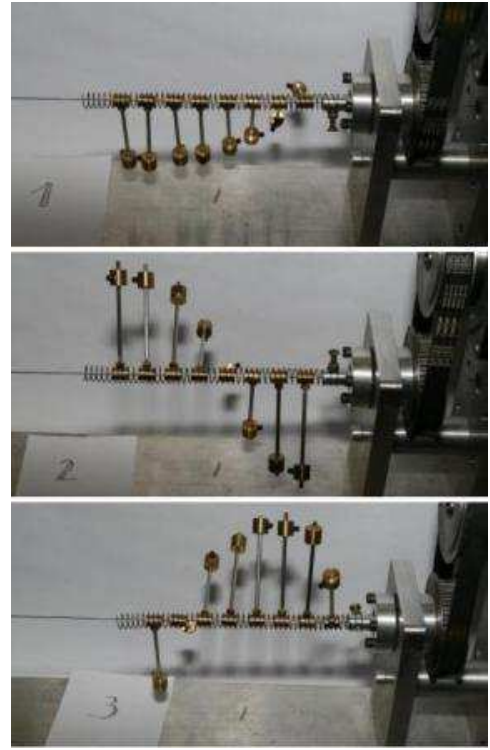,width=0.85\linewidth}
\caption{Pictures of three stationary states of the pendula chain obtained for
one single given driving amplitude and frequency. The upper graph corresponds to
the lowest energy state and the lower one describes the largest energy regime.
An approximate energy hierarchy is 1:10:100.} 
\label{fig:chain}\end{figure}

We report here the discovery of a third stationary state which can be
qualitatively understood as the motion back and forth of a kink-like structure
in the short pendula chain. The new stationary regime apears to be completely
different from the two cases considered earlier e.g. in
\cite{RJ1,RJ2,RJ3,RJ4,RJ5}. As a matter of fact, such a dynamics creates a new
frequency in the system and furnishes a tool to divide the input frequency by
odd fractions (we shall illustrate chain end oscillations with frequency
$\Omega/3$ or $\Omega/5$, where $\Omega$ is the driver frequency). The value of
the odd divider depends both on the input frequency range and on the length of
the chain. Let us recall that the previously discovered  two regimes are
\textit{synchronized} to the driver (same input and output frequencies).

\section{Model Equations.}

The dynamics of the chain of $N$ pendula is naturally described by
the Frenkel-Kontorova model  \cite{kivsharbook} 
\begin{equation}\label{discr-SG}
\ddot u_n+\delta\dot u_n-\sigma^2\ (u_{n+1}+u_{n-1}-2u_n)+
\omega_0^2\ \sin u_n=0,
\end{equation}
where overdot means derivation with respect to time. The variable $u_n$ is the
angular deviation of the n$^{th}$ pendulum, $\omega_0$ is the eigenfrequency of
a single pendulum and $\sigma$ is proportional to the linear torsion constant of
the spring (for our experimental chain $\omega_0=15.1\,$Hz, $\sigma=32.4\,$Hz).
The damping coefficient $\delta$ is phenomenological, it has been
evaluated in the experiments as approximately $\delta=0.01\omega_0$. 
This is the value actually used in the numerical simulations. The applied
periodic driving is here modeled by the boundary conditions
\begin{equation}\label{bounds}
 u_0(t)=b\,\cos(\Omega t),\quad u_{N+1}=u_N
\end{equation}
which model a forced end in $n=0$ and a free other end in $n=N$.

It is worth insisting on the fact that the chain is submitted to a
\textit{prescribed boundary value} (the datum of $u_0(t)$), not to a given
force acting on the first particle. In the experimental setup, the motion of
the virtual pendulum $u_0$ is the driving engine motion which has indeed a
prescribed motion obtained through feedback control. This has an important
fundamental consequence: the energy absorbed by damping is continuoulsly
compensated by the action of the driving and therefore the motion of the chain
is quite similar to that of an undamped device (for which, in a stationary
regime, the driving would not give energy to the chain).

The experiment consists thus in driving the short chain pictured in
Fig.\ref{fig:chain} with a frequency in the forbidden band gap
($\Omega<\omega_0$), which actually does not excite linear modes. Without
external perturbation the system locks to a periodic solution with low output
amplitude $u_N(t)$. Depending on the value of an external kick one makes the
system bifurcate to two different stationary states and we thus obtain a
tristable behavior (with approximate hierarchy of energies 1:10:100). The reader
will find on the web page \cite{web-site} a movie of the experiment where the
system is first set in the high-energy new stationary stable regime and then put
down successively to the two others stable states by taking energy off. 

\begin{figure}[ht]
\epsfig{file=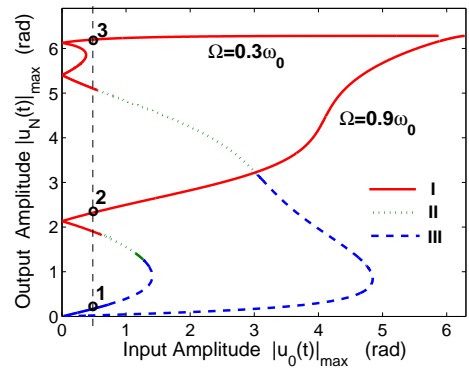,width=0.8\linewidth} 
\caption{Analytic input-output amplitude dependences for different oscillation
frequencies of stationary states given by formulas \eqref{solA} where continuous
line represents solution (I), dotted line solution (II) and dashed line solution
(III), plotted for two frequencies, namely $0.9\,\omega_0$ and $0.3\,\omega_0$
as indicated. The points 1, 2 and 3 corresponds to the stable regimes with a
single driving amplitude $|u_0(t)|_ {max}=0.5\,$rad. The points 1 and 2
represent the situations when the whole chain oscillates with the driving
amplitude $0.9\,\omega_0$ but with different output amplitudes. The point 3
corresponds to the driving frequency $0.3\,\omega_0$ and describes kink motion
forth and back. As the experiments and numerical simulations show (and this is a
main finding of the paper), the latter regime can also be reached with a driving
frequency $0.9\,\omega_0$ three times larger than the one actually used.}
\label{fig:analytics} \end{figure}

To develop an analytical description of the process,
let us consider the continuous approximation of eq.(\ref{discr-SG}) by
substitutions $t\rightarrow\omega_0 t$, $n=\omega_0
x/\sigma$. Neglecting dissipation we obtain the sine-Gordon equation
\begin{equation}\label{SG}
x\in[0,L]\ :\ u_{tt}-u_{xx}+\sin u =0,
\end{equation}
where $L=N\sigma /\omega_0$. The mixed Dirichlet and Neumann Boundary Conditions
$u(0,t)= b\,\sin (\Omega t)$ (driven boundary), $u_x(L,t)=0$ (free end boundary)
allows to seek the following periodic stationary solutions \cite{RJ5} 
\begin{equation}\label{sol11}
u(x,t)=4\arctan\left[\sqrt{\left|\frac{r s }{b}\right|}
{\cal X}(x){\cal T}(t)\right], 
\end{equation}
where one has three choices ($\cn$, $\sn$ and $\dn$ are the standard Jacobi
elliptic functions)
\begin{eqnarray}\label{solA}
&{\rm (I)}& {\cal X}=\cn(\beta (x-L),\mu), \quad {\cal T}=\cn(\omega t,\nu),
\nonumber \\
&{\rm (II)}& {\cal X}=\dn(\beta(x-L),\mu), \quad {\cal T}=\sn(\omega t,\nu),\\ 
&{\rm (III)}& {\cal X}=\dn(\beta (x-L)+ {\mathbb K}(\mu ),\mu ), \quad 
{\cal T}=\sn(\omega t,\nu ). \nonumber
\end{eqnarray}
Here ${\mathbb K}(\mu )$ stands for a complete elliptic integral of the first
kind of modulus $\mu $. These families of solutions are parametrized by the two
free constants $\omega$ and $\nu\in[0,1]$, then for solutions of type (I) the
remaining parameters are given by
\begin{align}
&b=\omega^4\nu ^2(1-\nu ^2), \quad s =\omega^2\nu ^2 \nonumber\\
&2r =1-\omega^2+2\omega^2\nu ^2+\sqrt{(1-\omega^2)^2+4\omega^2\nu ^2}, 
\nonumber\\
&\beta^2=\frac{b+r ^2}{r }, \quad \mu ^2=\frac{r ^2}{b+r ^2}, 
 \label{bb}
\end{align}
while in both cases (II) and (III) they read
\begin{align}
& b=\omega^4\nu ^2,\quad s =-\omega^2\nu ^2, \nonumber\\
&2r =1-\omega^2(1+\nu ^2)+\sqrt{[1-\omega^2(1+\nu ^2)]^2-4\omega^4\nu ^2},
\nonumber\\ 
&\beta^2=r , \quad \mu ^2=1-\frac{b}{r ^2}. 
\label{kkpp1}
\end{align}
Note that $r $ should be real valued and positive which may restrict the allowed
values of $\omega$.

\begin{figure}[ht]
\epsfig{file=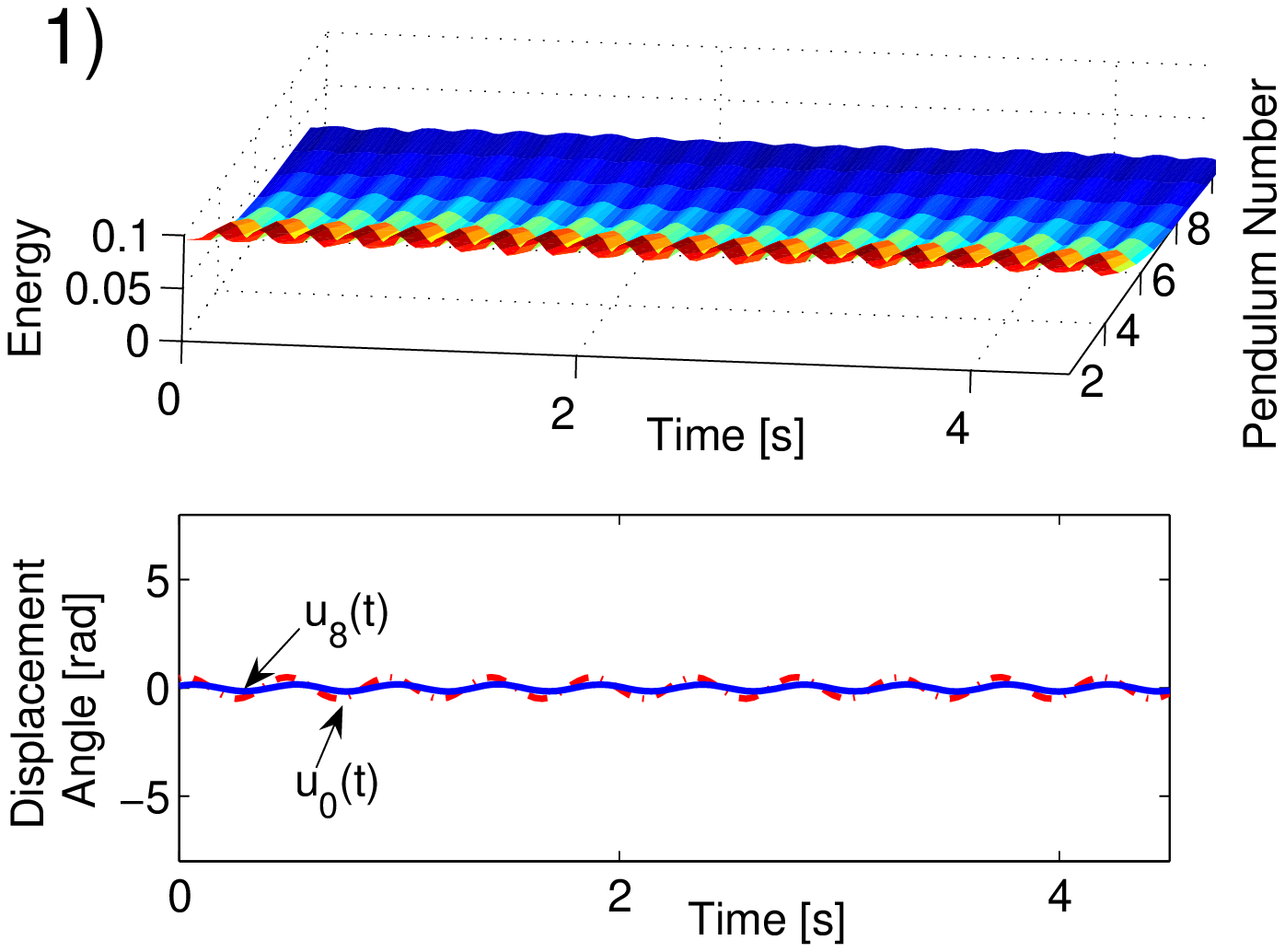,width=0.8\linewidth} 
\epsfig{file=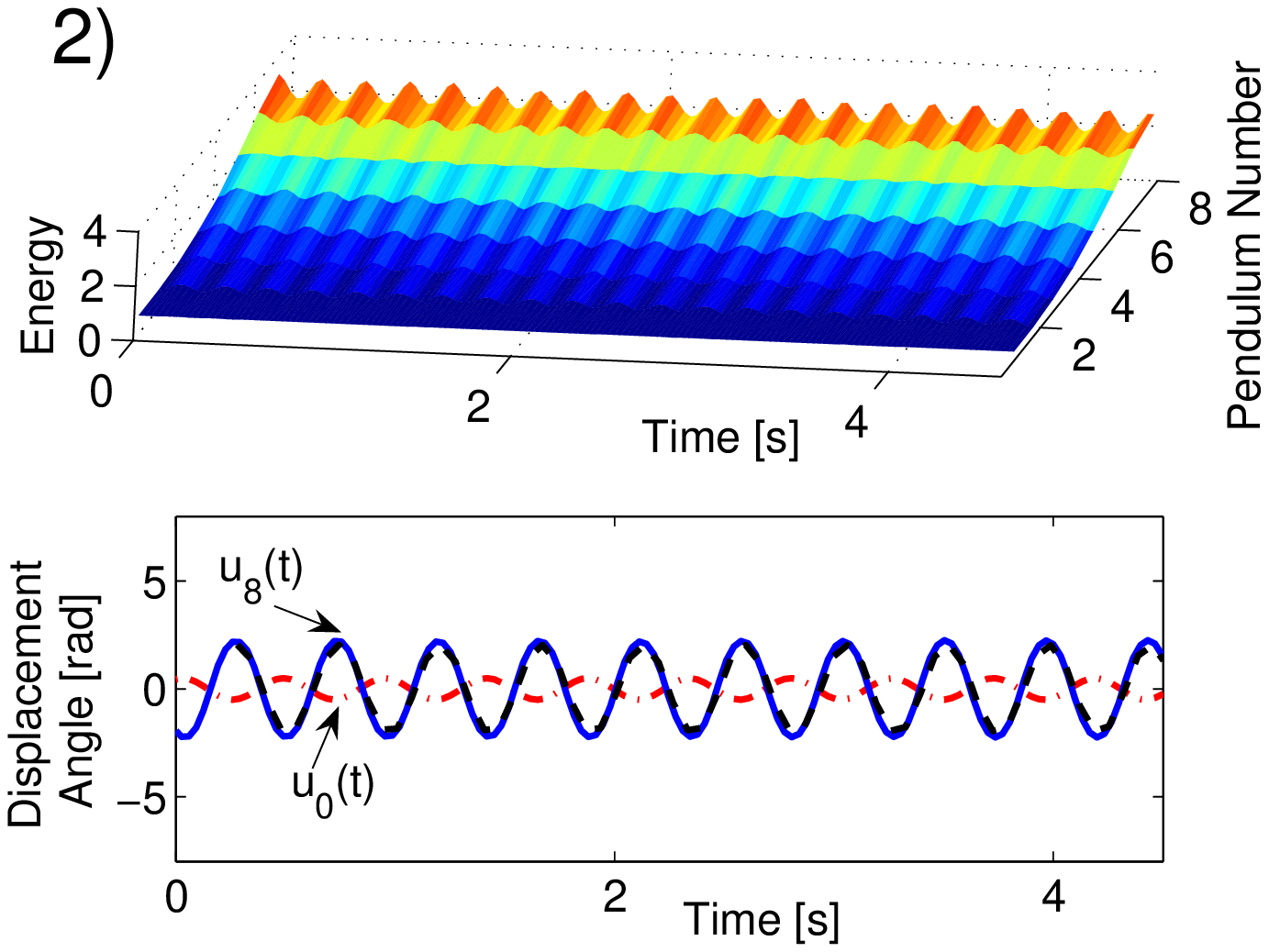,width=0.8\linewidth}
\epsfig{file=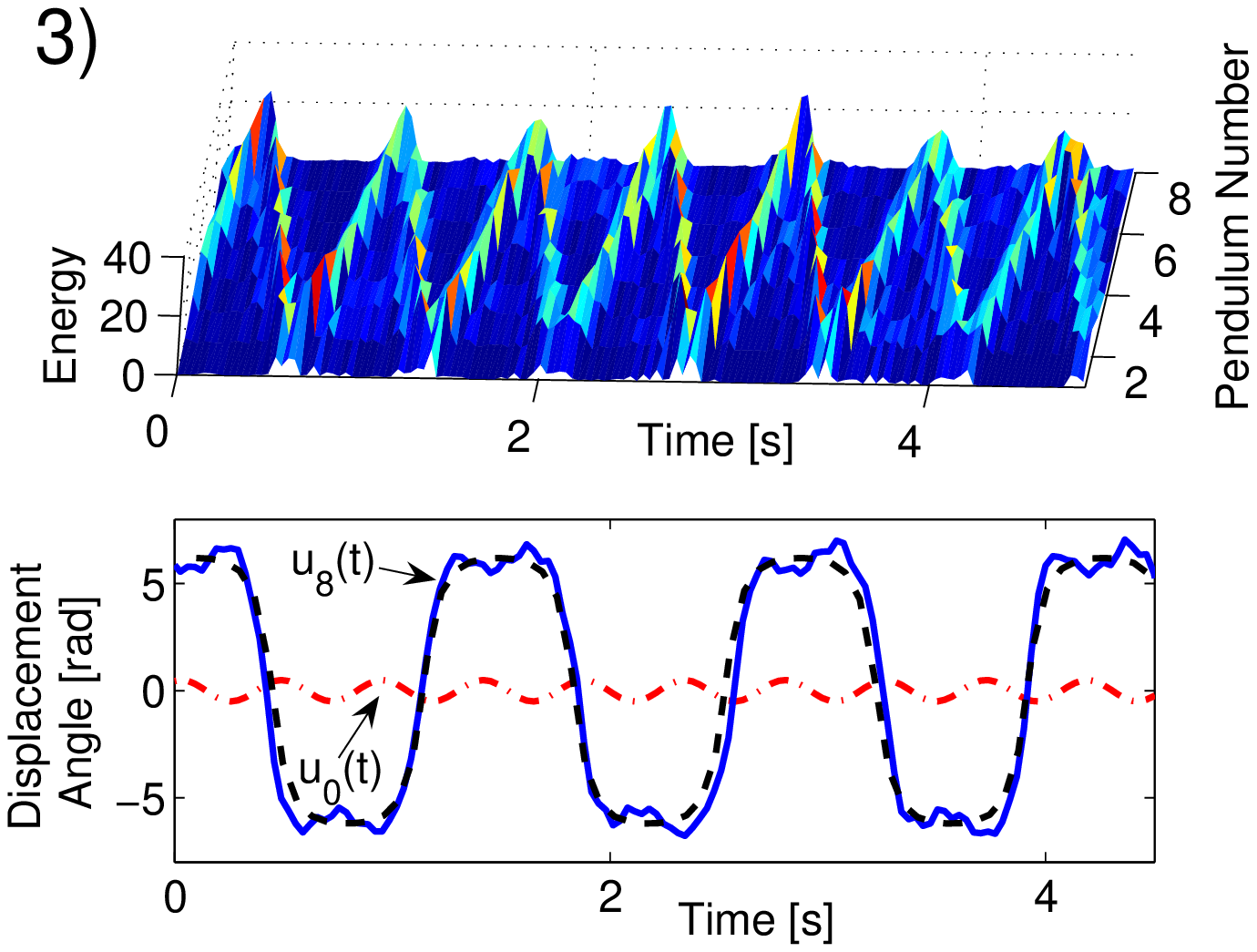,width=0.8\linewidth}
\caption{Numerical simulations on Frenkel-Kontorova model \eqref{discr-SG} with
a damping constant $\delta=0.01\,\omega_0$  and 8 pendula. The time evolution of
pendula energy and input-output oscillations are displayed corresponding to the
points 1), 2) and 3) in Fig. 2. The driving amplitude is
$|u_0(t)|_{max}=0.5\,$rad and its frequency $\Omega=0.9\,\omega_0$ for all 3
cases. This results in the same output frequency oscillations $\Omega$ in graphs
1) and 2) but $\Omega/3$ output oscillations in graph 3). Dashed lines display
analytical curves obtained from \eqref{solA}, while dotted-dashed and solid lines represent time evolution of input and output oscillations, respectively.}\label{fig:simuls1}\end{figure}

\section{Tri-stability and Frequency Division.}

Since the experiments (confirmed by numerical simulations later on) show
that the frequency $\Omega/3$ can also be excited, we assume that the period of
the time dependent part ${\cal T}(t)$ of the stationary solutions \eqref{solA}
coincide with an odd integer fractions of the driving frequency $\Omega$.
Recalling that the period of ${\cal T}(t)$ is $4{\mathbb K}(\nu)/\omega$ we
require thus
\begin{equation}\label{KK}
\omega=2\Omega{\mathbb K}(\nu )/(m\pi). 
\end{equation}
where $m$ is an odd integer. For a given value of the parameter $\nu\in[0,1]$,
the above relation fixes the second parameter $\omega$ in terms of the driving
frequency $\Omega$. Therefore fixing $\Omega$ (driver frequency) and varying
$\nu$ one can plot the output amplitude $u(N,t)$ in terms of the input
$u(0,t)$ from the analytic expressions (\ref{solA}).
We display this dependence for $\Omega=0.9$ (in units of
$\omega_0$) as a the full line in fig.\ref{fig:analytics} where different colors
report to different solutions. We also plot (dashed line) the
output amplitude for a driving frequency $\Omega=0.3$. Therefore, to the given
driver amplitude $\max_t |u(0,t)|=0.5$ may correspond two stable synchronized
states (points 1 and 2 on the graph) having the driver frequency $0.9$ and one
more stable state with frequency $0.3$. 

It is then a simple matter to check that the stationary state related to
point 3 of the plot of fig.\ref{fig:analytics} corresponds effectively
to our numerical simulations, and hence to the experiments of
fig.\ref{fig:chain}. It is done in fig.\ref{fig:simuls1} where the last plot
shows the result of a numerical simulation (full line) compared with the
analytic solution (dahsed line) related to point 3 of fig.\ref{fig:analytics}.
Wa have also plotted the time evolution of the total energy of each pendulum
given by
\begin{align}\label{energy}
E_n=&\frac{1}{2}\dot u_n^2+
\frac{\sigma^2}{4}\left[(u_{n+1}-u_n)^2+(u_{n-1} -u_n)^2\right]\nonumber\\
&+\omega_0^2(1-\cos u_n).
\end{align}
The numerical simulations of the process are done by applying to the model 
(\ref{discr-SG}) the boundary conditions (\ref{bounds})
with  $b=0.5$ and $\omega=0.9\,\omega_0$, together with an initial condition
where a few pendula at the end of the chain are given large initial amplitude.
For instance, to reach the new stationary state (3) of fig.\ref{fig:simuls1},
the chosen initial amplitude is $2\pi$, while for the value $\pi$, the system
locks to the state (2).

It is worth noting that both experiments (as those displayed in \cite{web-site})
and numerical simulations contain intrinsic damping. Still the analytic
solutions of the continuous undamped sine-Gordon model fit strikingly well
numerical simulations of the discrete damped Frenkel-Kontorova model
(\ref{discr-SG}). This is a general property of such \textit{short length driven
systems} to lock on fundamental solutions of the undamped limit, as previously
displayed in \cite{RJ3,RJ4,electric}. The main fact is that, without damping,
the chain in a stationary regime does not absorb energy and the boundary value
does not transfer any power to the chain. With damping, the driving boundary
does transmit power to the chain in such a way as to compensate exactly the
losses. The point is that the system is submitted to a \textit{prescribed
boundary value} which adapts to the amount of lost power and keeps amplitude and
frequency constants (in the experiments presented in fig.\ref{fig:chain}, the
engine has a feedback driving mechanism that controls the amplitude and
frequency). Last, the proof that the analytical solutions constitute an attractor for the damped system is, as far as we know, an open question.

\begin{figure}[ht]
\epsfig{file=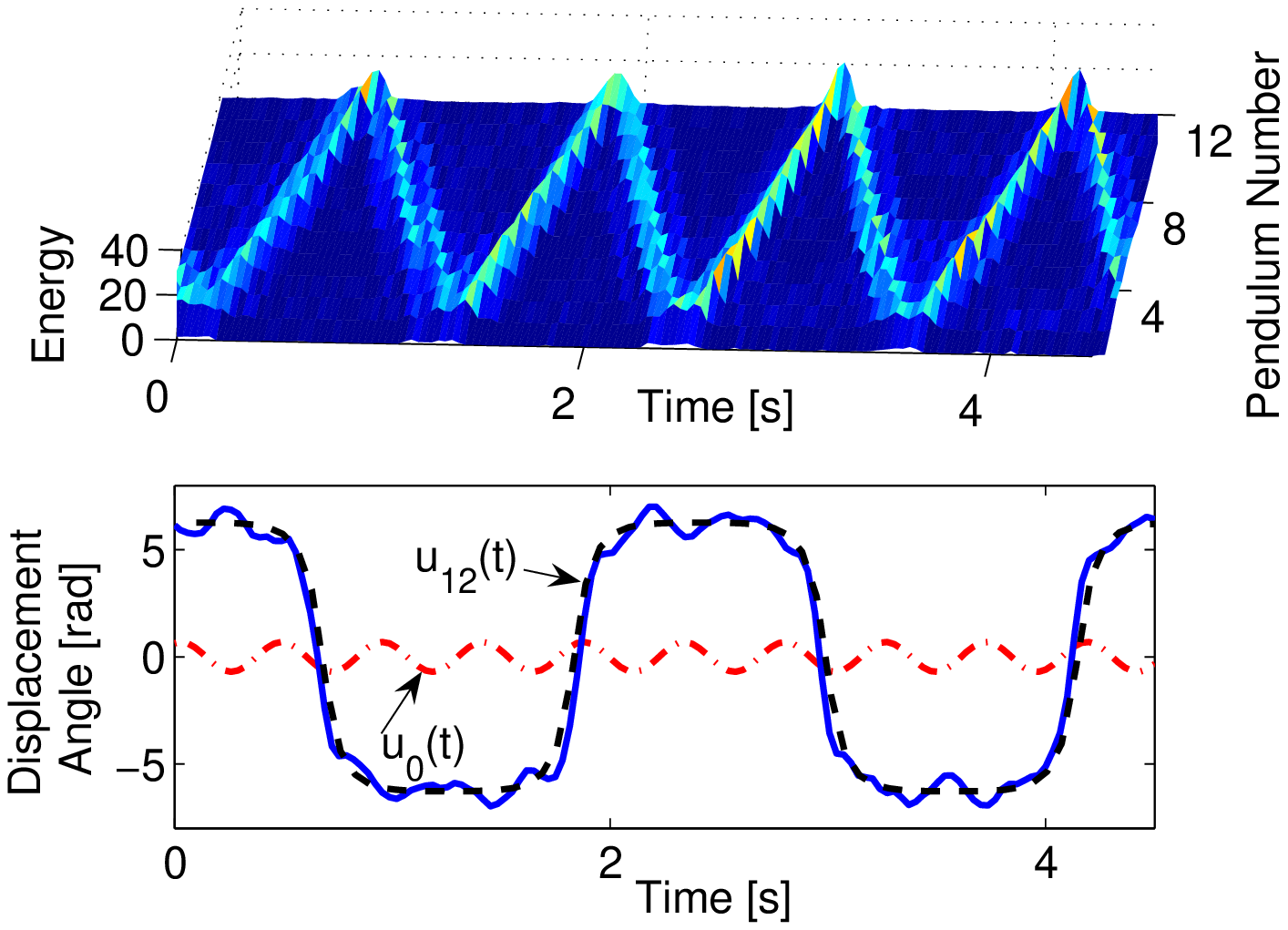,width=0.8\linewidth} 
\epsfig{file=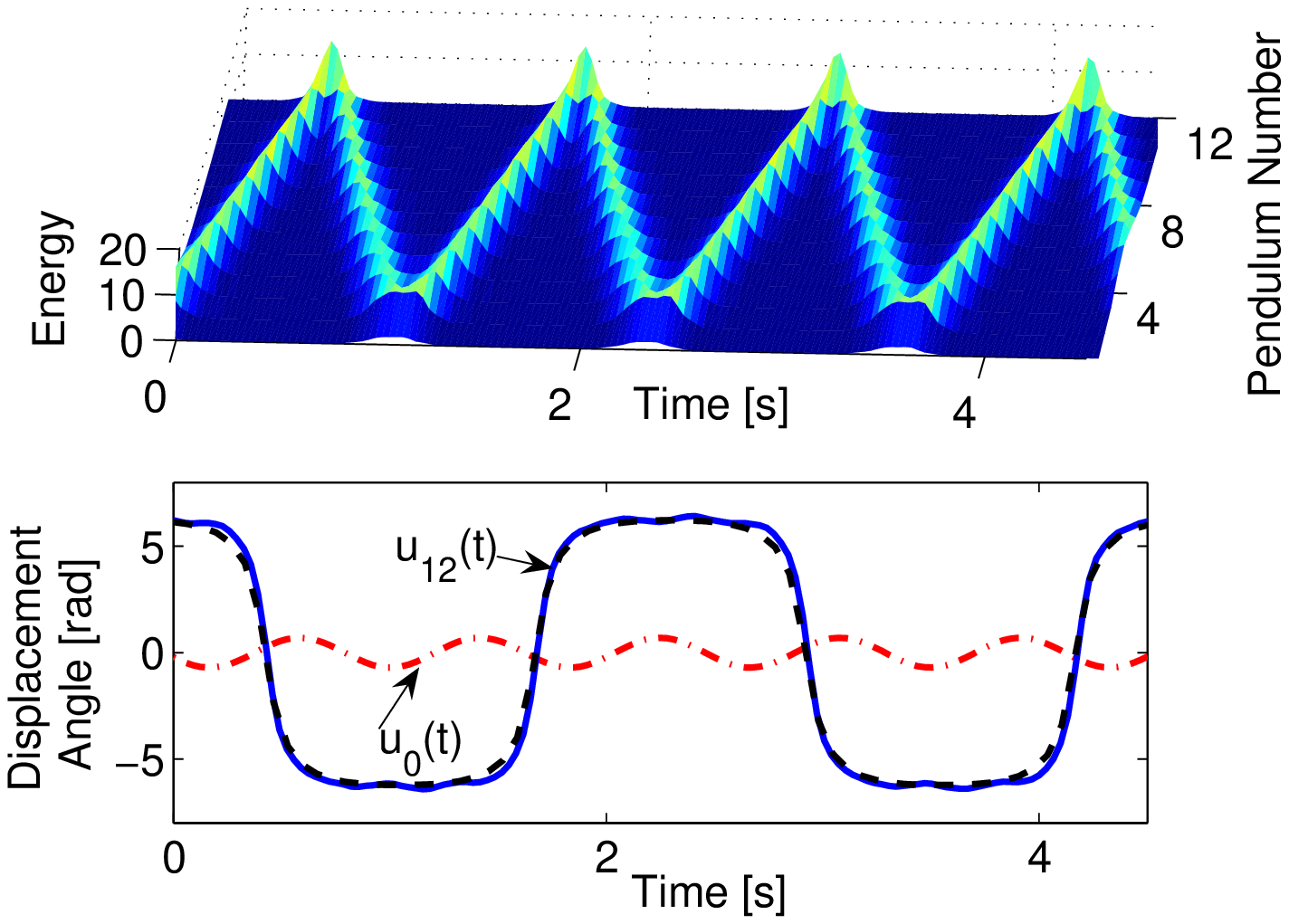,width=0.8\linewidth}
\caption{Time evolution of pendula energy and input-output oscillations for the
chain consisting of 12 pendula. As seen one gets the frequency division on 5 at
the output with respect to the input frequency when the input frequency is
$\Omega=0.9\,\omega_0$ (upper graph). In the lower graph the driving frequency
is $\Omega=0.5\,\omega_0$ and one has frequency division by factor
3.\label{fig:long}} \end{figure}

Thus we have actually demonstrated the possibility of conceiving a
\textit{frequency divider} with which the driving frequency can be divided by
3, 5, 7, depending on the chain length. For example the fig.\ref{fig:long} 
shows numerical simulations on a chain of 12 pendula with a resulting frequency
division by 5 at a driving frequency $0.9\,\omega_0$ and by 3 at a driving
frequency $0.5\,\omega_0$. In such a case we have obtained that the
frequency is divided by 5 if the driving frequency $\Omega$ is in the range
$0.88\omega_0<\Omega<0.92\omega_0$, while the same chain can divide the
frequency by 3 when $0.41\omega_0<\Omega<0.6\omega_0$.

\section{Conclusion.}

 For a single monochromatic driving (fixed amplitude and frequency), we have
demonstrated experimentally and numerically the existence three states which
have been given analytic expressions (in the continuous limit): the first one is
the quasi-linear solution (actually a breather-like tail), the second one
resembles half a \textit{breather}, both of them oscillating with the driving
frequency $\Omega$, and which were already known as the building blocks of
nonlinear bistability. The discovered third state resembles a \textit{kink}
moving back and forth with the frequency $\Omega/3$. These sates have been given
explicit analytical expressions in the continuous limit: the first state is
described by the solution of type (III) in (\ref{solA}) while the solution of
type (I) describes altogether the \textit{``half-breather''} with frequency
$\Omega$ and the \textit{``oscillating kink''} with frequency $\Omega/3$. 

The process of frequency division is thus induced by the motion back and forth
of a kink-like structure inside the chain. It is possible to extend these
studies to other realistic physical systems governed by the sine-Gordon
equation. We expect such a new stationary stable regime to be interesting
for applications where one is interested in producing an odd fraction of the
driving  frequency. The fraction number depends both on the length of the chain
and on the input frequency range. 

Last but not least, many other well known nonlinear systems exhibit nonlinear
bistable behavior, as e.g. the nonlinear Schr\"odinger equation or the coupled
mode system in Bragg media, and this discovery is very likely to apply also
there. 

\section{Acknowledgements.} We thank Dominique Chevriaux for the production 
of the movie on the pendula chain. R. Kh. acknowledges financial support of the
Georgian National Science Foundation (Grant No GNSF/STO7/4-197) and USA Civilian
Research and Development Foundation (award No GEP2-2848-TB-06).

\end{document}